# Beyond 100 THz-spanning ultraviolet frequency combs in a non-centrosymmetric crystalline waveguide

Xianwen Liu[1], Alexander W. Bruch[1], Juanjuan Lu[1], Zheng Gong[1], Joshua B. Surya[1], Liang Zhang[2], Junxi Wang[2], Jianchang Yan[2] & Hong X. Tang[1]

Ultraviolet frequency combs enable applications ranging from precision spectroscopy to atomic clocks by addressing electronic transitions of atoms and molecules. Access to ultraviolet light via integrated nonlinear optics is usually hampered by the strong material dispersion and large waveguide attention in ultraviolet regions. Here we demonstrate a simple route to chip-scale ultraviolet comb generators, simultaneously showing a gap-free frequency span of 128 terahertz and high conversion efficiency. This process relies on adiabatic quadratic frequency translation of a near-visible supercontinuum sourced by an ultrafast fiber laser. The simultaneous cubic and quadratic nonlinear processes are implemented in single-crystalline aluminum nitride thin films, where chirp-modulated taper waveguides are patterned to ensure a broad phase matching. The heterodyne characterization suggests that both the near-visible and ultraviolet supercontinuum combs maintain high coherence. Our approach is also adaptable to other non-centrosymmetric photonic platforms for ultrafast nonlinear optics with scalable bandwidth.

[1] Department of Electrical Engineering, Yale University, New Haven, CT 06511, USA. [2] R & D Center for Semiconductor Lighting, Institute of Semiconductors, Chinese Academy of Sciences, Beijing 100083, China. Correspondence and requests for materials should be addressed to H.X.T. (email: hong.tang@yale.edu)

Optical frequency combs that transmit equally-spaced, mutually-coherent spectral lines have revolutionized the field of ultrafast science and metrology[1]. Extending this technique into the ultraviolet (UV) region gives direct access to the electronic transitions of atoms and molecules, making it desirable for a myriad of applications. A exemplary case is precision spectroscopic sensing for chemical detection[2,3] and trace gas monitoring[4] by mapping their UV absorbance. The broad UV spectra also enrich the trapping and cooling of atom and ion species for atomic clocks[5–7] and quantum memories[8,9]. Conventionally, UV frequency combs are achieved from mode-locked lasers by leveraging nonlinear frequency conversion including high-harmonic generation in atomic gases[10], cascaded harmonics in bulk nonlinear crystals[11], and supercontinua in microstructure optical fibers[12,13]. While these schemes permit spectral transfer into deep-UV regimes, it requires bulk optics and intense pump energies, which are unfavorable for out-of-the-lab applications.

Nanophotonic architectures enable tight light confinement for enhanced optical nonlinearities as well as readily engineered dispersion, and have been exploited as a viable route towards four-wave mixing based microcombs with low power threshold and high scalability[14]. Attempts to transfer such microcombs into UV regimes are usually hampered by the strong normal material dispersion and large waveguide attenuation therein. Photonic chip-based supercontinua are equally important for serving as ultra-broadband comb sources in various photonic platforms[15–20], where soliton-induced dispersive waves are employed for coherent spectral transfer into visible (VIS) and mid-infrared regions. Nonetheless, UV supercontinuum microcombs are still in its infancy, and have only been demonstrated in silica-based waveguides[21] owing to its small material dispersion and flexible waveguide fabrication for tailoring a zero-integrated dispersion in the UV region.

Cascaded harmonic generation, on the other hand, provides an alternative route for spectral transfer to the UV wavelengths. By engineering the quasi-phase matching (QPM) in periodically poled lithium niobate waveguides[22] and microresonators[23], efficient UV harmonics below 400 nm have been accessed. Yet, the operating bandwidth of such QPW structures is commonly limited (~10 THz) for either pulsed[23] or continuous-wave (cw) pump[24]. In contrast, adiabatic frequency conversion is known for broadband operation[25], and has primarily been employed for sum-frequency (SF) generation in aperiodically poled bulk nonlinear crystals[26–28]. On-chip implementation of adiabatic SF transfer processes are recently theoretically discussed in linear-tapered waveguides for enhanced efficiency and bandwidth[29].

In this work, we exploit adiabatic frequency conversion using chirp-modulated taper waveguides and construct a UV microcomb with broad and gap-free envelopes. This process relies on adiabatic second-harmonic (SH) and SF translation of a broadened near-VIS supercontinuum in normal group-velocity dispersion (GVD) regimes. The simultaneous quadratic $\chi^{(2)}$ and cubic $\chi^{(3)}$ nonlinear processes are performed in non-centrosymmetric aluminum nitride (AlN) thin films. The resultant UV microcomb not only maintains a high degree of coherence but also exhibits a large frequency span of ~128 THz (namely 360–425 nm) and high off-chip efficiency of 6.6%, which is applicable for monitoring trace gases (for example HONO and $NO_2$ at 360–380 nm[4]) and interfacing with atoms and ions (for example $^{171}$Yb at 399 nm[6] and $^{40}Ca^{+}$ at 397 nm[7]). Our approach relaxes the bandwidth-efficiency trade off and is promising for on-chip ultrafast nonlinear optics.

## Results

**Concept and design**. Figure 1a schematically illustrates the principle of the UV microcomb generator: a near-VIS seed pulse around 780 nm is initially broadened via a $\chi^{(3)}$ process and simultaneously translated into the UV band via broad SH and SF conversion. This physical process is feasible in a photonic platform such as AlN by leveraging its intrinsic $\chi^{(2)}$ and $\chi^{(3)}$ susceptibilities. The large transparency window (above 200 nm) of AlN also makes it robust to the optical damage induced by multi-photon absorption[30]. The AlN film is epitaxially grown on *c*-plane (0001) sapphire substrate by metal-organic chemical vapor deposition, and has been developed for integrated nonlinear optics from near-VIS to infrared regions[31–35]. Based on a single-crystalline AlN thin film, we have demonstrated a low-loss (~8 dB cm$^{-1}$) UV microring resonator[36], suggesting its capability for integrated UV photonics.

Since wurtzite AlN belongs to $C_{6v}$ (6 mm) symmetry, incident light with a vertically-polarized electric field will experience a much larger $\chi^{(2)}$ susceptibility (above 8.6 pm V$^{-1}$ at 780 nm[33,37]) than that of horizontally-polarized light. Therefore, we engineer modal phase matching between fundamental and second-order transverse magnetic ($TM_{00}$ and $TM_{20}$) modes (Fig. 1b) for securing high SH and SF efficiency. Note that the phase-matching wavelengths ($\lambda$, $TM_{00}$ mode) are highly sensitive to the waveguide width ($w$), namely $d\lambda/dw \approx 0.46$, indicating a narrow phase-matching window of a uniform waveguide.

Alternatively, by leveraging adiabatic taper structures[29], it is possible to construct an efficient, broadband $\chi^{(2)}$ converter for the conceived UV microcomb in Fig. 1a. Such broad $\chi^{(2)}$ processes are enabled by adiabatically tuning the wavevector mismatch $\Delta k$ ($\lambda$, $z$) cross the phase-matching locations $z_0$ for each interacting wavelength component[26], which means

$$\Delta k(\lambda, z) \approx \Delta k(\lambda, z_0) + \Delta z \frac{\partial(\Delta k)}{\partial z}\Big|_{z=z_0} \quad (1)$$

Here $\Delta k(\lambda, z_0) = 0$ denotes the perfect phase-matching condition, while the phase factor $\Delta k(\lambda, z)\cdot\Delta z \approx 0$ is ensured by the extremely small modulus of $\partial(\Delta k)/\partial z$ at a moderate interaction length $\Delta z = z - z_0$. Meanwhile, the adiabatic interaction relaxes the temporal walkoff induced by the group-velocity mismatch of interacting fundamental and SH modes in nanophotonic waveguides (see Supplementary Note 1).

In practice, it is challenging to achieve quasi-continuous $\chi^{(2)}$ frequency translation from conventional linear-tapered waveguides because of the limited lithography precision (shown later). Meanwhile, the degraded near-VIS peak power along the waveguide will hamper the SH and SF efficiency in the latter of the linear-tapered waveguide. To address both issues, we exploit a chirp-modulated taper waveguide, which concatenates multiple adiabatic but length-distinct linear-taper segments (Fig. 1c). Therefore, the phase-matched frequency components will be slightly varied in each segment and obtain the opportunity to experience high near-VIS peak power at the former of the waveguides.

The GVD and nonlinearity of one of the phase-matched AlN waveguides are depicted in Fig. 1d. Owing to the dominant normal material dispersion in the near-VIS band, the waveguides naturally exhibit a normal GVD. The nanoscale waveguide cross section also yields an extraordinary nonlinearity ($\gamma \approx 9.5$ W$^{-1}$ m$^{-1}$ at 780 nm), at least a 40-fold improvement over that in silica-based waveguides ($\gamma \approx 0.24$ W$^{-1}$ m$^{-1}$)[21,38]. As a result, a coupled pulse energy of 237 pJ is sufficient to enable a near-VIS supercontinuum broadening from 600 to 1050 nm as predicted in Fig. 1e using the generalized nonlinear Schrödinger equation (see the "Methods" section). Such near-VIS spectra primarily arise from self-phase modulation (SPM) and are slightly modified around the spectral wings due to the occurrence of optical wave breaking (OWB)[39]. The corresponding temporal evolution is shown in Fig. 1f, confirming the degraded pulse peak power along the waveguide as mentioned earlier. The obtained near-VIS

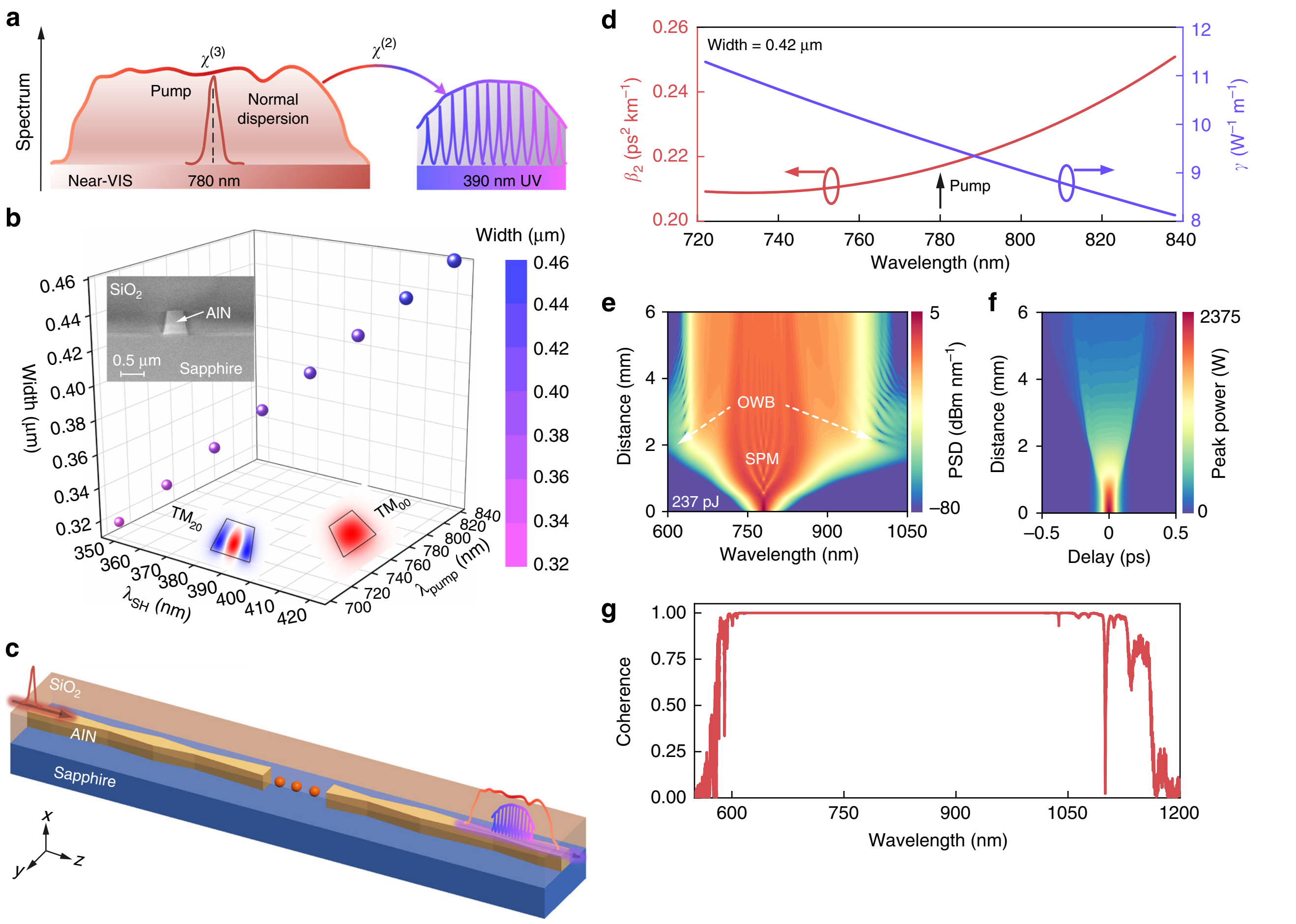

**Fig. 1** Principle of ultraviolet microcomb generators. **a** Schematics of spectral translation to the UV regime (purple curve) from a broadened near-VIS supercontinuum (red curve) in normal GVD regimes. **b** Calculated phase matching of near-VIS $TM_{00}$ and UV $TM_{20}$ modes (electric fields in the lower insets) in straight AlN waveguides using a finite-element method (FEM) solver. The right color bar denotes the varied waveguide widths. Upper inset: a scanning electron microscope (SEM) image of a cross-sectional AlN waveguide capped with silicon dioxide ($SiO_2$), showing a fixed height of 0.5 μm. **c** Illustration of the chirp-modulated taper waveguide based on single-crystalline AlN thin films. The red solid spheres indicate omitted waveguide segments and *x*, *y*, *z* are the waveguide coordinates. **d** Calculated second-order GVD $\beta_2$ (left) and nonlinearity coefficient $\gamma$ (right) of a phase-matched AlN waveguide (cross section: 0.42 × 0.5 μm$^2$) by FEM. Here $\gamma = 2\pi n_2/(\lambda A_{\rm eff})$ is estimated from the effective modal area $A_{\rm eff}$ and the nonlinear refractive index $n_2 = 3 \times 10^{-19}$ m$^2$W$^{-1}$[31]. Spectral (**e**) and temporal (**f**) evolution of the near-VIS pulse in the chirp-modulated taper AlN waveguide accounting for the distance-dependent parameters (see Methods). Similar results are observed for straight and linear-tapered waveguides. The incident pulse is around 780 nm with an on-chip energy of ~237 pJ (average power $P_{\rm ave}$ of 19 mW), and the right color bars denote the on-chip power spectral density (PSD) and peak power, respectively. **g** Calculated modulus of the first-order coherence for the near-VIS supercontinuum in (**e**) at a distance of 6 mm (see the "Methods" section)

spectrum is also robust to the input pulse noise and inherits a high degree of coherence from the seed laser, as suggested in Fig. 1g with the calculated modulus of the first-order coherence $\left|g_{12}^{(1)}\right| \approx 1$ over the entire bandwidth (see the "Methods" section). Therefore, through adiabatic SH and SF translation of the near-VIS supercontinuum comb, coherent UV microcombs are anticipated.

**Phase-matched UV supercontinuum combs**. The engineered AlN waveguides are patterned from an 0.5-μm-thick epitaxial film with optimized electron-beam lithography and dry etching processes as detailed in refs. [33,36]. Subsequently, the chip is capped with $SiO_2$ and cleaved to expose the waveguide facets. An example of the waveguide cross section is presented in the inset of Fig. 1b. The experimental setup is sketched in Fig. 2a. The pump source operating around 780 nm is delivered by a turn-key femtosecond (fs) laser (Toptica FemtoFiber pro NIR) with adjustable average power up to ~140 mW at a pulse duration of ~100 fs (full width at half maximum) and a repetition rate ($f_{\rm rep}$) of ~80 MHz. To leverage the large $\chi^{(2)}$ susceptibility of AlN, the electric field of the incident pulse is selected to be vertically polarized by employing a half-wave plate such that the $TM_{00}$ mode inside the waveguide is excited. The spectra of outgoing lights are then captured by a grating-based optical spectrum analyzer (OSA, Ando AQ6315E, 350–1750 nm). The details are presented in Methods.

To unveil the principle of the UV microcomb via adiabatic SH and SF conversion, we have prepared three distinct AlN waveguides (straight, linear-tapered, and chirp-tapered geometries) to probe the underlying phase-matching behavior. For consistence, all the waveguides possess 5-mm-long phase-matching portion (widths of 0.3–0.5 μm), while the total waveguide length is 6 mm including the wide taper couplers at the end facets (widths of 3 μm) for improved light coupling efficiency. An example of the overall geometry is illustrated in Supplementary Note 2.

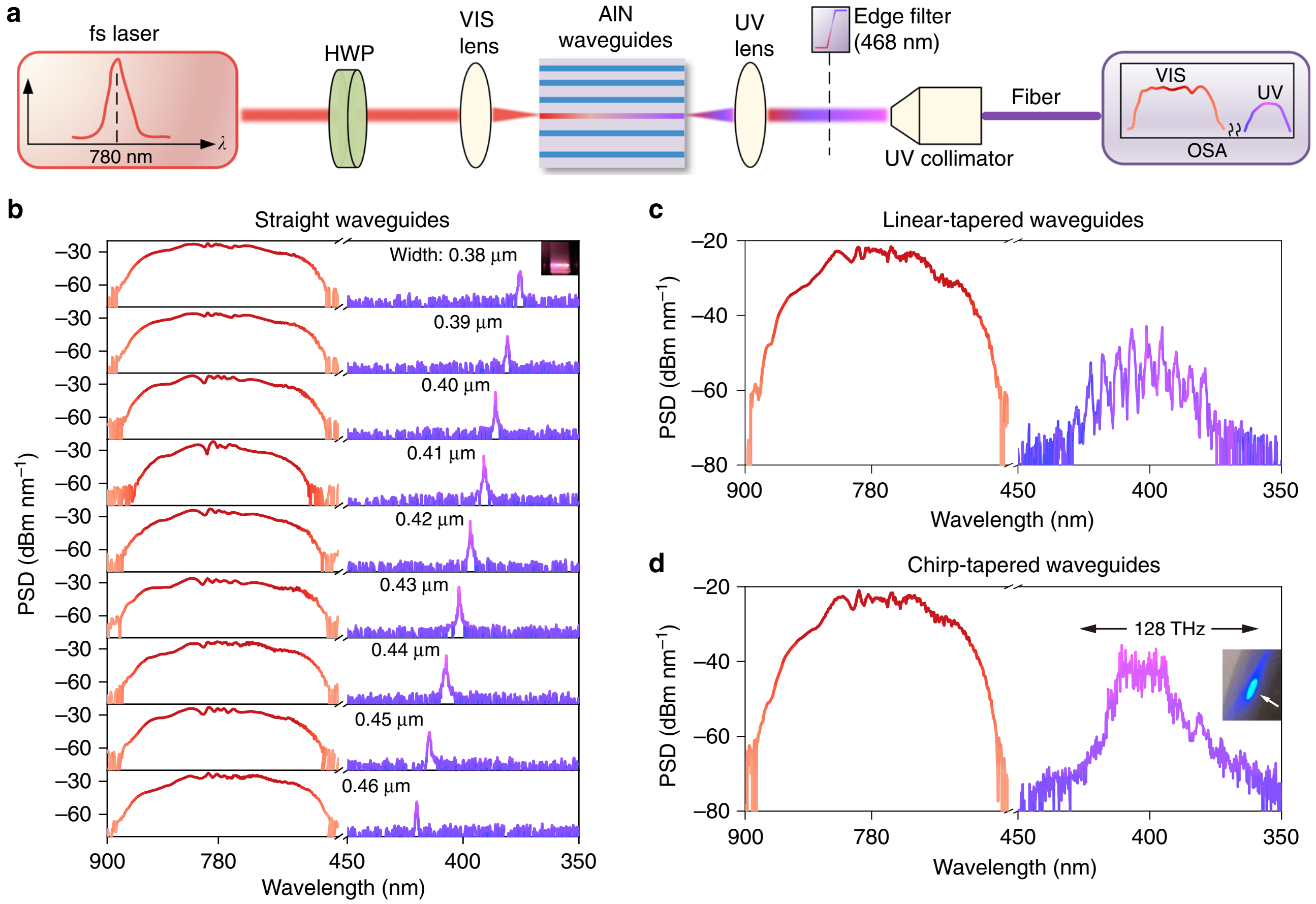

**Fig. 2** Phase-matched ultraviolet microcombs. **a** Sketch of the experimental setup. fs: femtosecond, HWP: half-wave plate, OSA: optical spectrum analyzer. To capture UV light, a short-pass edge filter is introduced to block concomitant near-VIS light (see Methods). **b** Near-VIS (red) and shifted UV (purple) spectra recorded from a set of straight AlN waveguides with varied widths (0.38–0.46 μm). Inset: a photograph of a glowing AlN chip upon UV light generation. **c** and **d** Near-VIS and UV Spectra respectively recorded from linear- and chirp-tapered AlN waveguides (tapering widths: 0.5–0.3 μm). The observable UV spectral span in (**d**) is ~128 THz. Inset of (**d**): a bright blue fluorescence image of collimated outgoing UV light (orientation denoted by a white arrow) on a piece of white paper. In (**b**–**d**), the on-chip pump pulse energy is ~237 pJ ($P_{ave}$ of 19 mW), and the wavelength resolution of the OSA is set to be 0.1 nm. The distinct noise floors of near-VIS and UV spectra arise from different monochromator modes selected

Figure 2b plots the optical spectra recorded from a set of straight waveguides (widths of 0.38–0.46 μm). At a coupled pump pulse energy of ~237 pJ ($P_{ave}$ of 19 mW), we observe smooth and flat-top supercontinuum spectra that are broadened from 650 to 900 nm. The on-chip spectral broadening (see Fig. 1e) could be much wider when considering the degraded efficiency (~10%) to deliver free-space near-VIS light into the optical fiber. The oscillatory central ripples and steep spectral edges are ascribed to the interplay between SPM and OWB, consistent with the simulation in Fig. 1e. Meanwhile, narrow-band UV comb peaks are observed in Fig. 2b, showing shifted wavelengths with the waveguide widths. The result correlates with the phase-matching design in Fig. 1b, suggesting spectral translation through SH and SF processes. The UV light is further verified to arise from the $TM_{20}$ mode by introducing an on-chip high-order mode filter into one side of the straight waveguide, where the resultant UV peaks exhibit 20 dB power variation upon feeding the incident pulse from different waveguide facets (see Supplementary Note 2). The signal-to-noise ratio of UV light in the OSA is up to 40 dB, and the free-space power could be much higher when accounting for the degraded efficiency (1.6%) for collecting free-space UV light into the optical fiber.

Guided by the shifted behavior of the UV peaks in Fig. 2b, we are able to expand the UV spectral bandwidth through adiabatic SH and SF translation using a single taper waveguide. According to the phase-matching design in Fig. 1b, we first linearly taper the waveguide width from 0.5 to 0.3 μm over a length of 5 mm (namely $|dw(z)/dz| = 4 \times 10^{-5}$). Followed by the adiabatic phase matching (see Supplementary Note 1), each interacting frequency component will be phase matched at different waveguide locations (namely $\Delta k(\lambda, z) \approx 0$), where adiabatic SH and SF conversion occurs rapidly in an effective adiabatic length. As plotted in Fig. 2c, based on the linear-tapered geometry, we indeed observe a broad UV spectrum with multiple peaks discretely spaced around the wavelength of 400 nm. The result is equivalent to combining the UV spectra in Fig. 2b, while the spectral gaps of neighboring UV peaks are ascribed to the limited lithography (beam step size of 2 nm, current of 1 nA) in producing the small width variation $|dw(z)/dz|$.

For practical applications, UV microcombs with a gap-free envelope are commonly preferred. For this purpose, we further employ a chirp-modulated taper waveguide concatenated with eight linear-tapered segments (see Fig. 1c, wide taper couplers being omitted for clarity). For each segment, the tapering width is within 0.5–0.3 μm while the tapering length is intentionally varied within 450–800 μm (total tapering length of 5 mm). Based on this architecture, the width variation $|dw(z)/dz|$ of each taper channel is slightly varied, which induces different phase matching behavior of adiabatic SH and SF processes, and thereby gap-free spectral translation. As displayed in Fig. 2d, by exploiting the chirp-tapered waveguide, the obtained UV comb exhibits a flat-top and smooth envelope, resembling the original near-VIS spectrum. The spectral flatness is not improved when we increase the number of taper segments to twelve. The observable spectral range in the OSA is ~360–425 nm, corresponding to a frequency span of ~128 THz. Such large $\chi^{(2)}$ upconversion bandwidth using adiabatic waveguides is on par with the result in adiabatic bulk nonlinear crystals[27], and represents a 10-fold improvement over that in QPM lithium niobate microresonators[23] and waveguides[24]. The realistic off-chip spectrum is even wider when accounting for 1.6% coupling efficiency of free-space UV light to the optical fiber. The $P_{\mathrm{ave}}$ of the UV light in the optical fiber is measured to be ~2.5 μW, consistent with the extracted $P_{\mathrm{ave}}$ of the UV spectrum in Fig. 2d using the expression: $P_{\mathrm{ave}} = \int \mathrm{PSD}(\lambda)\mathrm{d}\lambda$ (see Supplementary Note 3). In the inset of Fig. 2d, the free-space UV light induces bright fluorescence on a piece of white paper, suggesting high off-chip power.

**Characterization of UV supercontinuum combs**. We further compare the total power and conversion efficiency of the generated UV light in straight, linear-tapered, and chirp-tapered waveguides, respectively. As plotted in Fig. 3a, the total near-VIS transmittance of the three distinct waveguides is nearly identical, without noticeable degradation of the light guiding performance for linear- and chirp-tapered configurations. The decreasing transmittance with the launched power might arise from the enhanced multi-photon absorption of the near-VIS light and the increasing energy transfer to the more lossy $\mathrm{TM}_{20}$ UV mode, which experiences significant propagation loss (above 16 dB cm$^{-1}$) in the phase-matched AlN waveguides[36].

Figure 3b, c respectively plot the captured average UV power $\left(P_{\mathrm{UV}}^{\mathrm{out}}\right)$ and external conversion efficiency $\eta$ versus the total output power $\left(P_{\mathrm{total}}^{\mathrm{out}}\right)$. Here we define $\eta = P_{\mathrm{UV}}^{\mathrm{out}}/\left(P_{\mathrm{total}}^{\mathrm{out}}\right)$ so as to reveal the actual spectral translation from the broadened near-VIS supercontinuum to the UV light. This definition is also adopted in silica-based UV supercontinuum generation[21]. Note that the three types of waveguides under investigation maintain similar performance in the low-pump regime, whereas the chirp-tapered waveguide begins to outperform the other two when further elevating the pump energy. This can be explained by the alleviation of the degraded near-VIS peak power with the distance (Fig. 1f) in chirp-tapered waveguides, thereby leading to enhanced total SH conversion efficiency. When the ultrafast laser operates with full power (namely on-chip energy of ~420 pJ), we have achieved a maximum $P_{\mathrm{UV}}^{\mathrm{out}}$ of 0.56 mW in chirp-tapered waveguides, corresponding to a $\eta$ of 6.6%. The result is roughly a 2-fold improvement over that of uniform and linear-tapered geometries. The distinct slopes of the derived $\eta$ in Fig. 3c suggests the trait of chirp-tapered geometries at high pump levels. The maximum $\eta$ achieved is on a par with that of dispersive wave-based UV supercontinua in silica waveguides[21], although the absolute $P_{\mathrm{UV}}^{\mathrm{out}}$ value is scaled down proportionally when accounting for the available pump energy in our experiment.

We further assess the coherence of both the near-VIS and UV microcombs in Fig. 2d by beating a selected comb line with a narrow-linewidth cw reference laser. As illustrated in Fig. 4a, the reference laser is sourced by a Ti-sapphire laser (M2 SolsTis, 700–1000 nm, linewidth <50 kHz), which also allows access to a reference UV laser through the SH process in the chirp-tapered AlN waveguide. Therefore, a high spatial modal overlap can be ensured between the reference UV laser and the UV comb as they both belong to the $\mathrm{TM}_{20}$ mode. For UV light characterization, a short-pass edge filter (<468 nm) is employed to block the near-VIS light.

According to the radio frequency characterization in Fig. 4b, c at a resolution bandwidth of 6.5 kHz, both the near-VIS and UV microcombs are found to possess regular comb teeth. The insets highlight the beatnotes around 80 MHz, suggesting that the near-VIS and UV microcombs maintain the same $f_{\mathrm{rep}}$ as a consequence of simultaneous SH and SF conversion[35,40]. Figure 4d, e depicts the heterodyne beatnotes within a frequency span of 100 MHz for the near-VIS and UV microcombs, respectively. The repetition rate beatnotes $f_{\mathrm{rep}}$ are located around 80 MHz, while the other two beatnotes ($f_{\mathrm{beat,1}}$ and $f_{\mathrm{beat,2}}$) arise from the beating between the Ti-sapphire (SH UV) laser and two neighboring near-VIS (UV) comb lines, showing a relationship of $f_{\mathrm{beat,1}} + f_{\mathrm{beat,2}} \approx 80$ MHz. The zoom-in near-VIS and UV beatnotes around $f_{\mathrm{beat,2}}$ are plotted in the insets, where a high signal-to-noise ratio above 20 dB and a short-term 3-dB linewidth below 10 kHz are observed. Our results suggest that both the near-VIS and UV supercontinuum combs are highly coherent[10,17], and no noticeable comb linewidth broadening is observed after the $\chi^{(2)}$ translation process.

## Discussion

We demonstrate a deterministic yet simple avenue to generate UV supercontinuum microcombs with a gap-free envelope

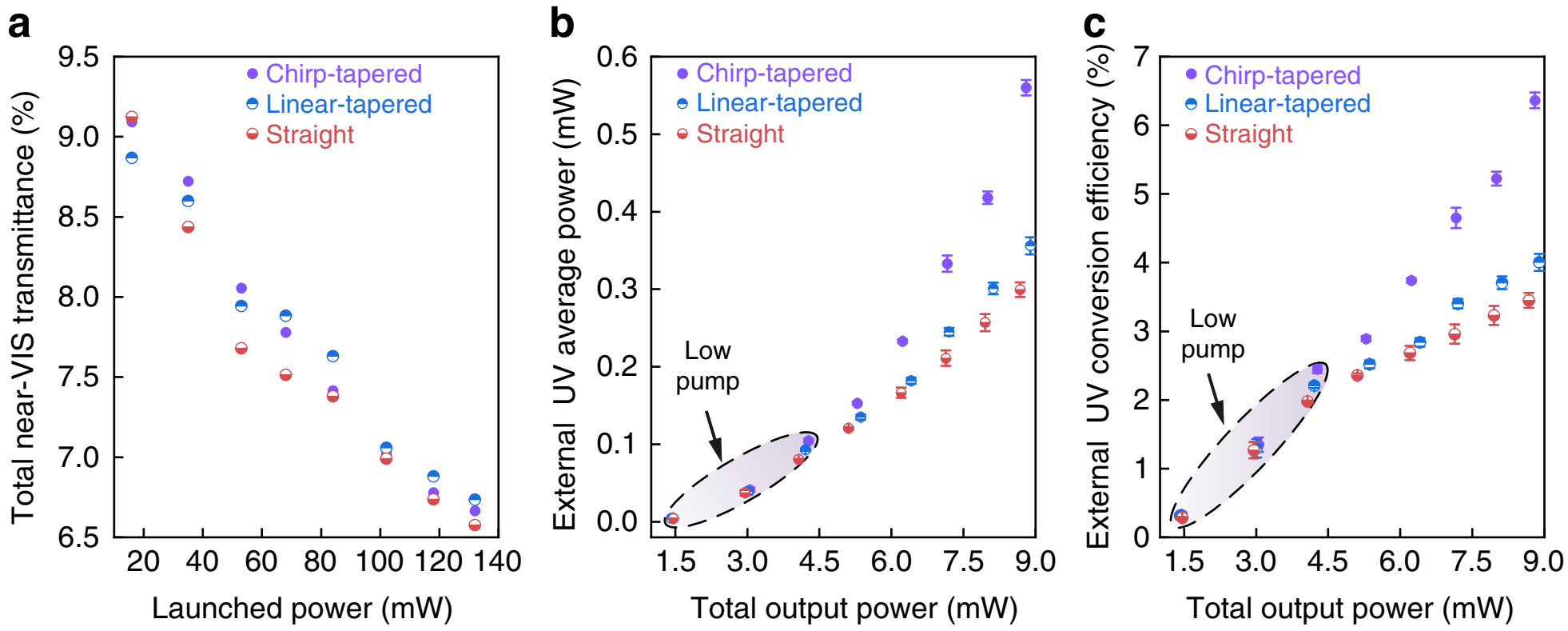

**Fig. 3** Efficiency characterization. **a**–**c** Performance comparison of chirp-tapered (purple), linear-tapered (blue), and straight (red) AlN waveguides with **a** total transmittance of near-VIS light, **b** collected average power of free-space UV light, and **c** external UV light conversion efficiency. The error bar describes the result variation recorded from other waveguides with the same geometry. In (**b**) and (**c**), the colored shadows reveal the low pump power applied

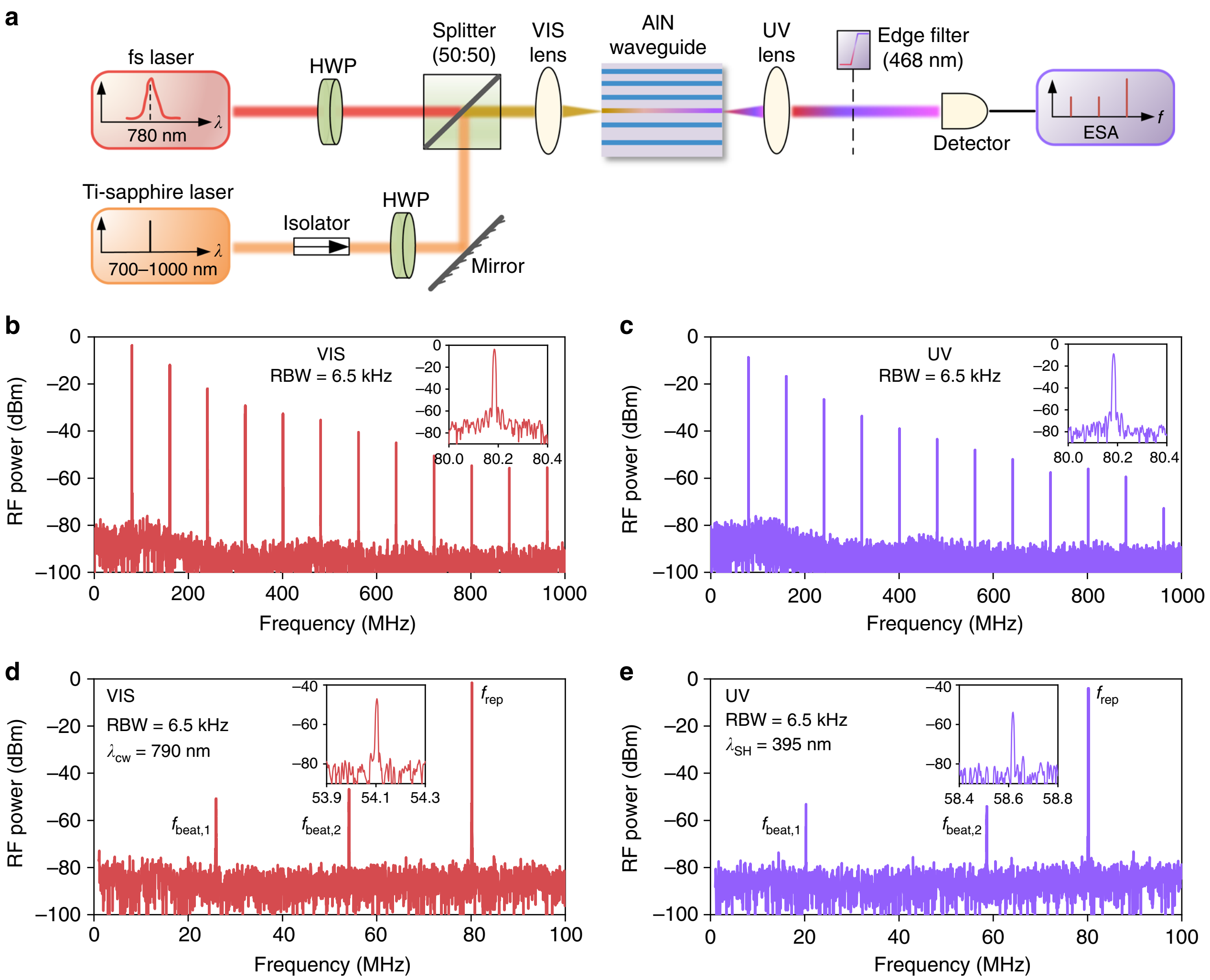

**Fig. 4** Coherence characterization. **a** Illustrated experimental setup for characterizing heterodyne beatnotes of near-VIS and UV supercontinuum microcombs. The fs and reference Ti-sapphire lasers are combined by a cube beamsplitter and delivered into the AlN chip. The outgoing lights are then sent into a low-noise UV-sensitive photodetector (Newport 1801, 320–1000 nm) followed by an electrical spectrum analyzer (ESA). **b** and **c** Radio-frequency (RF) spectra of free-space near-VIS (attenuated) and UV combs at a resolution bandwidth (RBW) of 6.5 kHz when the Ti-sapphire laser is off. The decreasing RF power with the frequency is induced by the limited detector bandwidth of 125 MHz. Insets in (**b**) and (**c**): zoom-in near-VIS and UV repetition rates around 80 MHz. **d** and **e** Heterodyne beatnotes of free-space near-VIS (attenuated) and UV microcombs (RBW of 6.5 kHz) when the fs and Ti-sapphire lasers are on. Here we lock the Ti-sapphire laser at $\lambda_{cw} = 790$ nm using a high-resolution wavemeter and obtain the SH UV laser at $\lambda_{SH} = 395$ nm with the free-space power above 10 µW. Insets in (**d**) and (**e**): zoom-in near-VIS and UV heterodyne beatnotes around $f_{beat,2}$ with short-term 3-dB linewidths below 10 kHz

spanning beyond 100 THz. This process is enabled by adiabatic SH and SF translation of a broadened near-VIS supercontinuum comb in a chirp-modulated taper AlN waveguide. The upconverison process not only maintains a high degree of coherence but also shows off-chip efficiency up to 6.6% (average power of 0.56 mW). Such coherent and efficient UV microcombs are feasible for on-chip UV spectroscopic sensing and atom-photonic interfaces. Our result suggests that chirp-modulated adiabatic waveguides are important media for ultrafast nonlinear optics with less susceptibility to the degraded pulse peak power with the propagation distance. The approach can be also adopted to other non-centrosymmetric waveguide materials such as lithium niobate[24], gallium nitride[41], and gallium phosphide[42] by taking full advantage of their intrinsic $\chi^{(2)}$ and $\chi^{(3)}$ nonlinearities.

## Methods

**Numerical simulation**. The evolution of the near-VIS pulse in chirp-modulated taper AlN waveguides is numerically investigated using the generalized nonlinear Schrödinger equation[43]:

$$\frac{\partial E(z,t)}{\partial z} = -\frac{\alpha}{2}E(z,t) + \sum_{k\geq 2} i^{k+1}\frac{\beta(z)_k}{k!}\frac{\partial^k E(z,t)}{\partial t^k} + i\gamma(z)\left(1+\frac{i}{\omega_0}\frac{\partial}{\partial t}\right)\left(|E(z,t)|^2 E(z,t)\right) \quad (2)$$

Here, $E(z, t)$ is the electric field of the optical pulse at the waveguide position $z$ and the delayed time $t$, $\omega_0$ is the center angular frequency of the incident pulse, and $\alpha = 140$ m$^{-1}$ (6 dB cm$^{-1}$) is the assumed linear waveguide loss around $\omega_0$. In the simulation, we also consider the influence of the distance-dependent GVD $\beta(z)_k$ and nonlinearity $\gamma(z)$ around $\omega_0$ as well as the adopted wide taper couplers at the wavegudie facets. The delayed Raman response is ignored in our simulation because no evident Raman effect is observed in the experiment. By normalizing $E(0, t)$ and its Fourier transform $\tilde{E}(0,f)$ to the launched pulse energy, the near-VIS power spectral density PSD($\lambda$) in wavelength scale is derived as:

$$\mathrm{PSD}(\lambda) = \frac{c}{\lambda^2} f_{rep}\left|\tilde{E}(z,f)\right|^2 \quad (3)$$

where $c$ is the light speed in vacuum.

The spectral coherence of the near-VIS spectrum is then numerically evaluated by calculating the modulus of $g_{12}^{(1)}(\lambda)$[44]:

$$g_{12}^{(1)}(\lambda) = \frac{<\tilde{E}_1^*(\lambda)\tilde{E}_2(\lambda)>}{\sqrt{<|\tilde{E}_1(\lambda)|^2><|\tilde{E}_2(\lambda)|^2>}} \quad (4)$$

Here $\tilde{E}_1(\lambda)$ and $\tilde{E}_2(\lambda)$ are wavelength-scale complex amplitudes of two independent near-VIS supercontinua, which are generated by imposing random quantum noise on the incident pulse. The ensemble average of the spectra is calculated from 100 shot-to-shot simulations with a frequency resolution of 20 GHz. The obtained $|g_{12}^{(1)}|$ versus the wavelength is plotted in Fig. 1g.

**Spectral measurement**. The aspheric UV lens (Thorlabs C610TME-A, numerical aperture NA = 0.6) and UV collimator (Thorlabs F671APC-405, NA = 0.6) for light collection in Figs. 2a and 4a are wavelength-dependent. To align the optical path for UV light, we place a short-pass edge filter (Semrock FF01-468/SP, transparency below 468 nm) after the UV lens to block the near-VIS light, which allows for UV signal detection using an ultra-sensitive spectrometer (JAZ-EL350, 350–1000 nm, Ocean Optics). After re-optimizing the alignment for UV light collimation, a grating-based OSA is employed for high-resolution spectral measurement while the near-VIS light is greatly suppressed. The average power of outgoing light from the UV lens and optical fiber are respectively recorded by two calibrated silicon power sensors (Thorlabs S130VC and S140C). The high launched power in Fig. 3a is detected by a thermal power sensor (Thorlabs S314C). The transmittance of delivering free-space near-VIS and UV lights into optical fibers is respectively measured to be 10% and 1.6%, which is responsible for the reduced PSD recorded by the OSA.

## Data availability

The data that support the findings of this study are available from the corresponding authors upon reasonable request.

## Acknowledgements

This work is supported by DARPA SCOUT (W31P4Q-15-1-0006). H.X.T. acknowledges support from DARPA's ACES programs as part of the Draper-NIST collaboration (HR0011-16-C-0118), an AFOSR MURI grant (FA9550-15-1-0029), a LPS/ARO grant (W911NF-14-1-0563), a NSF EFRI grant (EFMA-1640959) and David and Lucile Packard Foundation. The authors thank Michael Power and Dr. Michael Rooks for assistance in the device fabrication, and Prof. Changling Zou at the University of Science and Technology of China for constructive discussion.

## Author contributions

X.L. and H.X.T. conceived the waveguide design. X.L. performed the device fabrication, measurement and numerical analysis with the assistance from A.B., J.L., Z.G. and J.S. L.Z., J.W. and J.Y. provided the AlN wafer. X.L. and H.X.T. wrote the manuscript with the input from all other authors. H.X.T. supervised the project.

## Additional information

**Supplementary Information** accompanies this paper at https://doi.org/10.1038/s41467-019-11034-x.

**Competing interests:** The authors declare no competing interests.

**Reprints and permission** information is available online at http://npg.nature.com/reprintsandpermissions/

**Peer review information**: *Nature Communication* would like to thank the anonymous reviewers for their contributions to the peer review of this work.

**Publisher's note:** Springer Nature remains neutral with regard to jurisdictional claims in published maps and institutional affiliations.